\documentclass[fontsize=11pt,a4paper]{scrartcl}

\addtokomafont{disposition}{\rmfamily}

\usepackage[T1]{fontenc}
\usepackage[utf8]{inputenc}

\usepackage{csquotes}
\usepackage[english]{babel}
\usepackage{hyperref}

\usepackage{graphicx}
\usepackage{authblk}
\usepackage[a4paper]{geometry}
\usepackage{lmodern}

\usepackage{color}

\ifdefined\REVIEW
   \linenumbers
   
\fi

\title{MaRDMO: Future Gateway to FAIR Mathematical Data}

\author[1]{Marco Reidelbach}
\affil[1]{Zuse Institute Berlin, Takustraße 7, 14195 Berlin, Germany}

\date{}

\begin{document}
\maketitle

\section*{ORCID}
Please list the ORCIDs of all contributing authors (if applicable).
\begin{itemize}
    \item Marco Reidelbach: \url{https://orcid.org/0000-0002-1919-1834}
\end{itemize}
%
%
\section*{Keywords}
Please provide 3 to 5 keywords that add context to your contribution. \\
\textbf{Keywords:} Mathematical Research Data, FAIR Data, Knowledge Graph
%
%
\section*{Abstract}
Mathematical research data plays a crucial role across scientific disciplines, yet its documentation and dissemination remain challenging due to the lack of standardized research data management practices. The MaRDMO Plugin addresses these challenges by integrating mathematical models, algorithms, and interdisciplinary workflows into the established framework of the Research Data Management Organiser (RDMO). Built on FAIR principles, MaRDMO enables structured documentation and retrieval of mathematical research data through guided questionnaires. It connects to multiple knowledge graphs, including MathModDB, MathAlgoDB, and the MaRDI Portal. Users can document and search for models, algorithms, and workflows via dynamic selection interfaces that also leverage other sources such as Wikidata. The plugin facilitates the export to the individual MaRDI services, ensuring data quality through automated validation. By embedding mathematical research data management into the widely adopted RDMO platform, MaRDMO represents a significant step toward making mathematical research data more findable, accessible, and reusable.

%
%
\section{Introduction}

Mathematics is often perceived as an abstract discipline, detached from the concept of research data and rarely producing it~\cite{Boege2023}. Many mathematicians still assert, “I have no data, I’m just doing math.” This perspective poses a challenge for research data management (RDM) at universities and research institutions. Anecdotal evidence also suggests that those responsible for supporting researchers in their RDM efforts often struggle with how to assist mathematicians in managing and documenting their research data. Unlike in other scientific disciplines, where standards are well established\footnote{\url{https://obofoundry.org/principles/fp-000-summary.html}; \textit{Last accessed on February 25th, 2025.}}, mathematics lacks these for its research data. In some cases, no concrete data management is expected at all\footnote{\url{https://www.math.harvard.edu/media/DataManagement.pdf}; \textit{Last accessed on February 25th, 2025}}. 

To address these needs, the Mathematical Research Data Initiative, MaRDI~\cite{MaRDI2022}, as part of the German National Research Data Infrastructure, NFDI~\cite{Hartl2021}, has taken significant steps toward defining RDM standards in mathematics. Recognizing the increasing complexity and interdisciplinary potential of mathematical research data, MaRDI has developed services\footnote{\url{https://portal.mardi4nfdi.de/wiki/MaRDI_Services}; \textit{Last accessed on February 25th, 2025.}} , guidelines~\cite{MaRDI2023}, and outreach material~\cite{Bacher2023} to support the mathematical research community. At the core of these efforts is the application of the FAIR (Findable, Accessible, Interoperable, and Reusable) principles~\cite{Wilkinson2016}, ensuring that mathematical research data can be systematically documented, shared, and reused. These efforts aim to establish RDM as an integral component of mathematical research, aligning with funding agency requirements\footnote{\url{https://rea.ec.europa.eu/open-science_en}; \textit{Last accessed on February 25th, 2025.}}.

Mathematical research data is not limited to mathematics but is widely used across scientific disciplines~\cite{MaRDI2022}. Mathematical models are essential for simulations in engineering and the natural sciences~\cite{Alobaid2022}, algorithms drive computational advances in fields such as medicine and economics~\cite{Albuquerque2025}, and interdisciplinary workflows support applications in the humanities~\cite{Kostre2022}. The increasing reliance on data-driven methods~\cite{Alexander2023} across all areas of research highlights the need for structured and standardized approaches to handling mathematical research data. Ensuring that mathematical research data adheres to FAIR principles is not only crucial for mathematics itself but also essential for achieving FAIR research data management in all other disciplines that rely on mathematical research data.

Algorithms, mathematical models, and interdisciplinary workflows are fundamental in most research disciplines. To manage these data types, MaRDI developed dedicated templates and ontologies. For example, the Mathematical Model Database (MathModDB) ontology~\cite{Schembera2023,Schembera2024,Schembera2025,Shehu2025} enables the standardized documentation of mathematical models, including their formulations, involved quantities, related tasks, and modeled problems across various fields. Similarly, the Mathematical Algorithm Database (MathAlgoDB) ontology~\cite{MathAlgoDB2022} enables the standardized documentation of algorithms, their corresponding problems, implementing software, and relevant benchmarks. At present, the corresponding knowledge graphs\footnote{\url{https://mtsr2024.m1.mardi.ovh/}; \textit{Last accessed on February 27th, 2025.}}\footnote{\url{https://mathalgodb.mardi4nfdi.de/}; \textit{Last accessed on February 27th, 2025.}} contain 123 documented models and 204 documented algorithms. For interdisciplinary workflows, MaRDI developed a standardized documentation template that captures mathematical models, process steps, methods, software, hardware, instruments, and datasets associated with specific research objectives~\cite{Boege2023}. These templates are integrated into the MaRDI Portal~\cite{Schubotz2023}, a wikibase envisioned as a future one-stop shop for mathematical research data.

Providing researchers from all disciplines with easy access to the aforementioned MaRDI services is essential for fostering the adoption of structured mathematical research data management. The MaRDMO Plugin\footnote{\url{https://github.com/MarcoReidelbach/MaRDMO-Plugin}; \textit{Last accessed on March 30th, 2025.}} serves as a crucial gateway by integrating these services into the Research Data Management Organiser, RDMO~\cite{Neuroth2018}, an established data management plan software widely used across disciplines~\cite{Enke2023}. Through guided interviews, the MaRDMO Plugin enables researchers to document their algorithms, mathematical models~\cite{Reidelbach2024}, and interdisciplinary workflows~\cite{Reidelbach2023} in a standardized form and publish them directly in the respective MaRDI services, ensuring efficient dissemination. Additionally, researchers can query these services to re-use existing mathematical models, algorithms, workflows, or parts thereof for their own research. By embedding MaRDMO within RDMO, institutions can seamlessly incorporate it into their existing infrastructure, allowing RDM staff—who are often already familiar with RDMO—to support mathematicians and researchers from other disciplines working with mathematical research data effectively. To further extend accessibility, MaRDI will also provide its own RDMO instance, provided through the basic services DMP4NFDI~\cite{Diederichs2024}, ensuring that researchers without institutional access to RDMO can still benefit from these services.

The remaining paper introduces the key functionalities of the MaRDMO Plugin, followed by a discussion of its future potential and development directions, with an emphasis on its role in advancing research data management in mathematics.

\section{The MaRDMO Plugin}\label{sec:the-mardmo-plugin}

In the following section, the MaRDMO Plugin is introduced as a tool for documenting and querying mathematical models, algorithms, and interdisciplinary workflows through distinct questionnaires. To assist researchers in their documentation efforts, the individual questionnaires are semi-automated and linked to several knowledge graphs. All this follows an underlying data model, which is introduced first. 

\subsection{Data Model}\label{sec:data-model}

Interdisciplinary research follows a structured process when addressing real-world problems, defining research objectives and employing various approaches to derive answers. These workflows can vary significantly depending on the research context. Some may be purely computational, where a problem is simulated based on a mathematical model and propagated using algorithms. Others may involve experimental components, where empirical data is collected, fed into a model, and subsequently analyzed through computational methods and some may follow data-driven or rule-based methodologies, such as statistical inference or heuristic optimization, without relying on an explicit mathematical model. Overall, however, data, methods (computational or experimental), mathematical models, instruments, software, and hardware are key aspects of these workflows but are not necessarily present in all cases.

MaRDI's interdisciplinary workflow documentation template structures these workflows into distinct processing steps, a concept closely related to Metadata4Ing~\cite{Iglezakis2023}, where input data is transformed into output data through computational or experimental methods. Each step is further associated with relevant tools, such as software or instruments, and is assigned to specific research fields or mathematical areas. This structured approach enables precise documentation of how individual data sets are modified and how they flow across disciplines.

\begin{figure}[ht]
    \centering
    \includegraphics[width=0.9\textwidth]{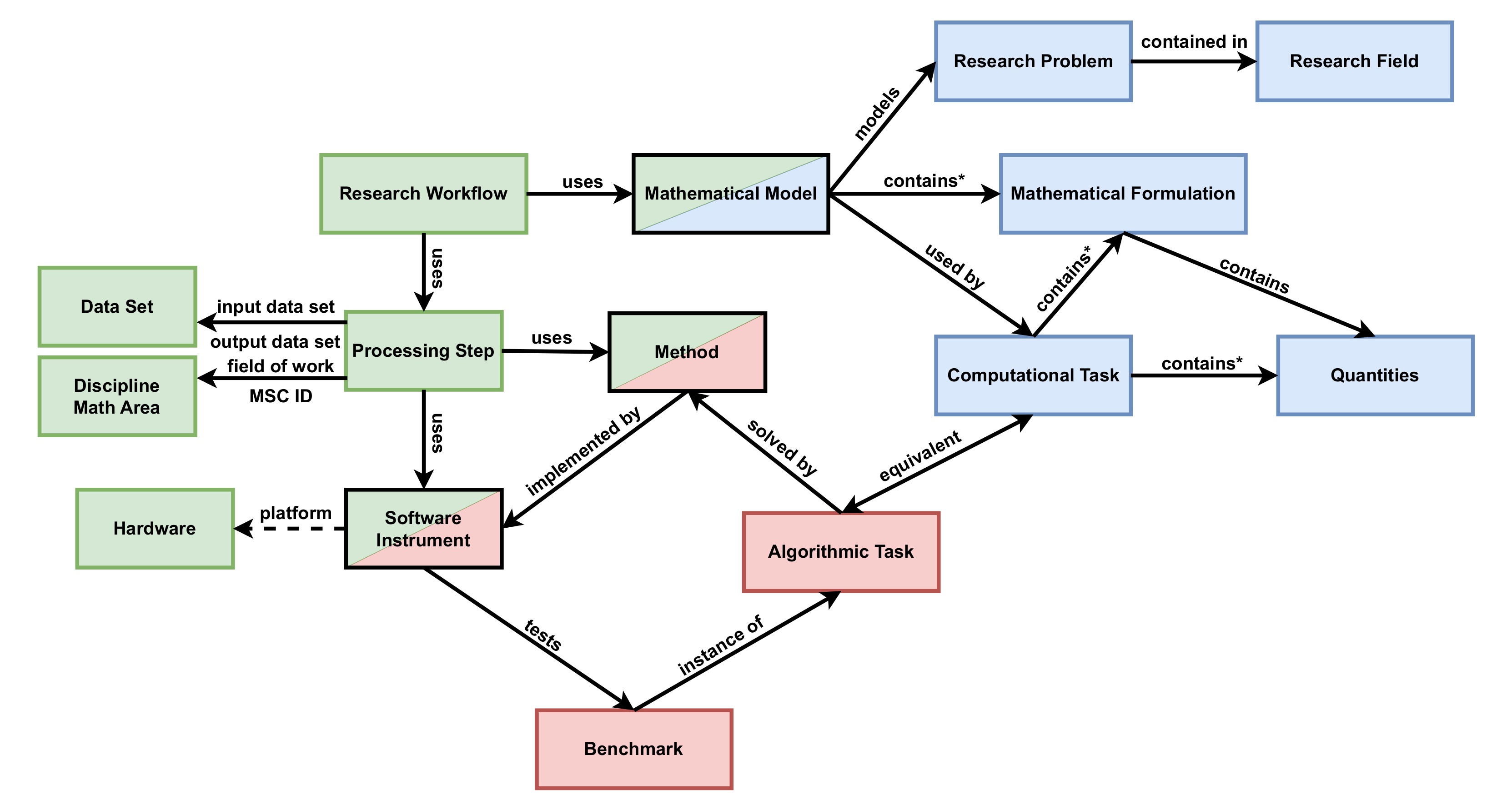}
    \caption{\small Data model employed by the MaRDMO Plugin. Nodes represent classes, while labeled, directed edges indicate relations between them. Classes relevant to algorithms, mathematical models, and interdisciplinary workflows are depicted in red, blue, and green, respectively. Two-colored nodes signify relevance to multiple data types. The \textit{Method} class is shared between interdisciplinary workflows and algorithms, but only methods classified as algorithms are considered part of the algorithmic domain. \textit{Software} and \textit{Instruments} are distinct classes, combined here to parallel the \textit{Tool} class in Metadata4Ing; however, only \textit{Software} belongs to the algorithmic domain and is linked to the \textit{Hardware} class. An asterisk in the edge label denotes the existence of multiple distinct relations. The publication class relevant for algorithms, mathematical models and interdisciplinary workflows is omitted for clarity.}
    \label{fig:figure1}
\end{figure}

To integrate interdisciplinary workflows into the MaRDI Portal, the MaRDMO Plugin leverages, where possible, existing items from both the MaRDI Portal and Wikidata~\cite{Vrandevic2014}. For example, existing items such as \textit{Research Workflow}\footnote{\url{https://www.wikidata.org/wiki/Q115682347}; \textit{Last accessed on March 7th, 2025.}} or \textit{Mathematical Model}\footnote{\url{https://www.wikidata.org/wiki/Q486902}; \textit{Last accessed on March 7th, 
2025.}} serve as classes in the data model for interdisciplinary workflows. Notably, an item describing a general \textit{Processing Step}\footnote{\url{https://portal.mardi4nfdi.de/wiki/Item:Q6486603}; \textit{Last accessed on March 20th, 2025.}} did not exist and was subsequently added to the MaRDI Portal. Instances of these classes are related by properties such as \textit{uses}\footnote{\url{https://www.wikidata.org/wiki/Property:P2283}; \textit{Last accessed on March 7th, 2025.}} or \textit{platform}\footnote{\url{https://www.wikidata.org/wiki/Property:P400}; \textit{Last accessed on March 7th, 2025.}}. Figure \ref{fig:figure1} (green boxes) provides a schematic representation of the integration of interdisciplinary workflows. Note that, for clarity, only the main aspects are shown. A complete list of employed items and properties can be found in the MaRDMO documentation\footnote{\url{https://portal.mardi4nfdi.de/wiki/MaRDMO}; \textit{Last accessed on March 20th, 2025.}}.

As discussed earlier, mathematical models form a crucial component of many interdisciplinary workflows, providing the theoretical foundation for computational analyses and simulations. To ensure proper documentation of these workflows, the MaRDMO Plugin integrates the MathModDB ontology. This enables a standardized description of models, including their formulations, quantities, related tasks, and modeled research problems across various fields (c.f. Fig. \ref{fig:figure1}, blue boxes).

Both workflows and models are connected to research fields and problems. While this might seem like a duplication, the associated fields and problems can differ. For instance, a workflow could focus on testing different computational methods for a particular model, making the distinction necessary for accurate documentation. 

MaRDMO further expanded to include the MathAlgoDB ontology, which provides a standardized description of algorithms (c.f. Figure \ref{fig:figure1}, red boxes). By linking algorithms to their implementing software, the algorithmic problems they address, and related benchmarks, MathAlgoDB enhances the overall documentation of the computational methods employed. Connecting these algorithms to the broader workflow context, MaRDMO now offers a more complete picture of the applied methods and their underlying computations. Furthermore, the computational tasks of MathModDB and the algorithmic problems of MathAlgoDB are interconnected, and MaRDMO can leverage this connection.

The MathModDB and MathAlgoDB ontology also support relations between items of the same class, not shown in Figure \ref{fig:figure1} but listed in the MaRDMO documentation, such as linking similar models or subclasses of algorithms. All these relations enhance the findability of algorithms, mathematical models and interdisciplinary workflows. 

By the time of this writing, interdisciplinary workflows, mathematical models, and algorithms are documented in distinct knowledge graphs. Federated SPARQL queries enable the implementation of the data model across the individual knowledge graphs. This setup, however, is likely to evolve in the future.

\subsection{Questionnaires and Functionalities in MaRDMO}\label{sec:questionnaires-and-funcionalities}

To facilitate the structured documentation of mathematical models, algorithms, and interdisciplinary workflows, the MaRDMO Plugin provides distinct questionnaires. These questionnaires guide researchers through the documentation process by collecting relevant metadata in a standardized format. Each questionnaire aligns closely with the corresponding ontology, ensuring consistency with MathModDB, MathAlgoDB, and the interdisciplinary workflow documentation template. The questionnaires vary in complexity, with different numbers of sections and questions depending on the type of research data being documented. Despite these differences, all questionnaires share a common structure that facilitates a streamlined and user-friendly documentation process.

\begin{figure}[ht]
    \centering
    \includegraphics[width=0.9\textwidth]{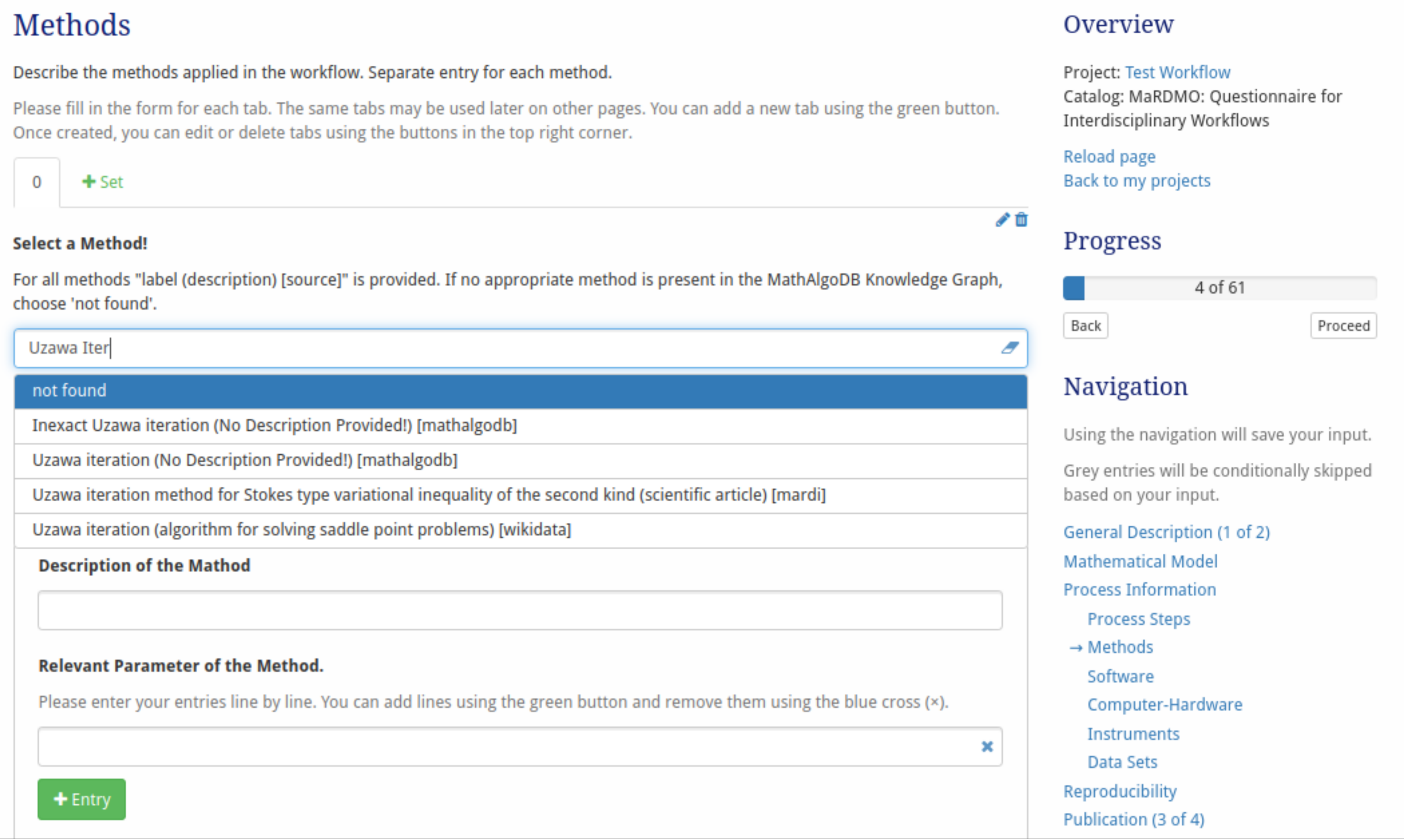}
    \caption{\small RDMO user interface with the \textit{Methods} page of the interdisciplinary workflow catalog. \textit{Uzawa Iteration} is searched, corresponding items are found in MathAlgoDB, the MaRDI Portal and Wikidata.}
    \label{fig:figure2}
\end{figure}

Each section of a questionnaire consists of one or more pages, which follow a structured format. The first step on either page involves selecting an appropriate item from an external source, e.g. for the documentation of a mathematical model, such an item could be the model itself, a research field, a research problem, a computational task, a quantity, or a mathematical formulation. For models, the preferred external source is MathModDB, while for algorithms, the primary reference is MathAlgoDB. If an item is not available in these sources, users can search alternative databases, including the MaRDI Portal or Wikidata. When an item from a lower-priority source is selected, MaRDMO automatically verifies whether it exists in a higher-priority source, ensuring consistency and minimizing redundancy. Item selection is supported by \textit{Dynamic Optionset Providers}, which query the respective sources via API or SPARQL, presenting results in an autocomplete fashion (c.f. Fig. \ref{fig:figure2}). Currently, most providers support a subset of four sources: MathModDB, MathAlgoDB, the MaRDI Portal, and Wikidata, but additional sources can be incorporated as needed.

Following the selection, the next step on the page presents an information box containing key details about the selected item. The content of this box depends on the item's class. For instance, a research field entry may include only the name and description, whereas a mathematical formulation entry contains additional details such as properties, formulas, symbols, and encoded quantities. If an existing item is chosen, MaRDMO queries the respective source for further details and automatically populates the information box. If no suitable item is found, users must manually provide the necessary information.

In addition to describing the selected item, the information box also displays connections to items of other classes, e.g a \textit{Research Problem} is contained in a \textit{Research Field}. As for the pure item information, described before, MaRDMO adds existing relations automatically to the questionnaire. If a relation to an item from another class is established, MaRDMO checks whether the corresponding item already exists in the respective section of the questionnaire, e.g. if the \textit{Research Field} containing the \textit{Research Problem} is already part of the \textit{Research Field} section. If it does, no further action is taken; otherwise, the related item is added to the appropriate section, and all available metadata is incorporated into the questionnaire. Dedicated automation schemes (c.f. Fig. \ref{fig:figure2}) define which data is added, how, and when, ensuring consistency and minimizing redundancy. This automation enhances efficiency and supports researchers by reducing manual input.

\begin{figure}[ht]
    \centering
    \includegraphics[width=0.9\textwidth]{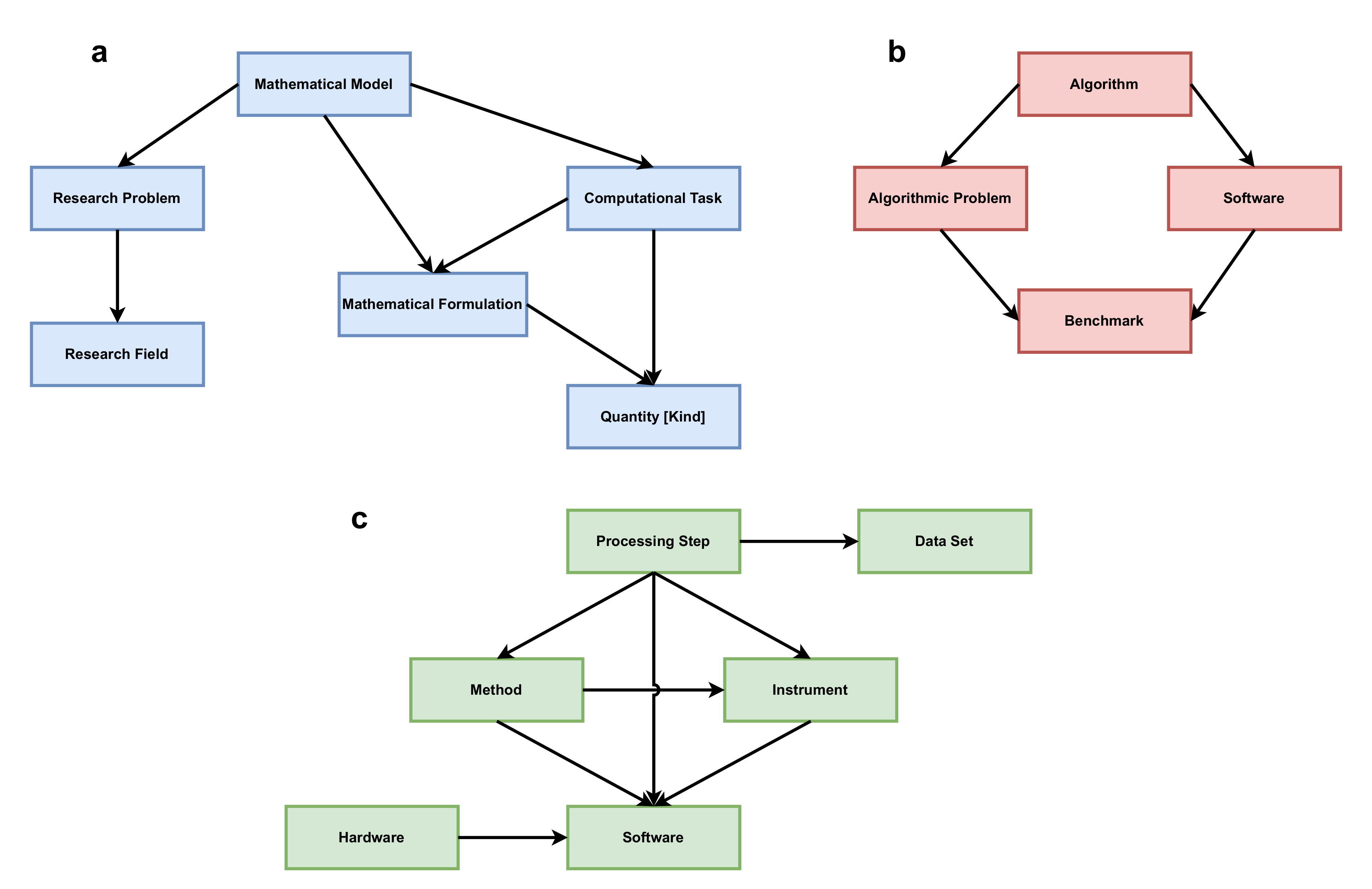}
    \caption{\small Automating schemes within the catalogs for mathematical models (a), algorithms (b) and interdisciplinary workflows (c) displaying the individual classes of the corresponding ontologies. The selection of items from either class leads to an automatic insertion of all downstream items. For clarity, the publication class, which is a downstream class of all other classes, has been omitted.}
    \label{fig:figure3}
\end{figure}

At the end of the pages, additional questions allow users to specify relationships between items of the same class. These include connections such as stating that one item \textit{generalizes}, \textit{approximates}, or \textit{is similar to} another. Again, MaRDMO pre-fills these fields where possible based on the information from the external sources. Since defining such relationships can be challenging for individual researchers these questions are optional and it is envisioned that domain experts will contribute to refining these connections through later curation efforts.

It is important to note that all three documentation catalogs share a common publication section. The main difference lies in how individual publications are linked to other items\footnote{\url{https://portal.mardi4nfdi.de/wiki/MaRDMO}; \textit{Last accessed on March 19th, 2025.}}. To accommodate this, MaRDMO provides a dedicated page in the questionnaire, allowing users to select existing publications from MathAlgoDB, MathModDB, the MaRDI Portal, or Wikidata. If an appropriate publication is found, its DOI or URL is added to the questionnaire. If no suitable publication is available in these sources, users can manually provide a DOI or URL. When a DOI is entered, MaRDMO automatically checks the four aforementioned sources, as well as the Crossref API\footnote{\url{https://api.crossref.org/works/}; \textit{Last accessed on March 19th, 2025.}}, the DataCite API\footnote{\url{https://api.datacite.org/dois/}; \textit{Last accessed on March 19th, 2025.}}, the DOI API\footnote{\url{https://citation.doi.org/metadata?doi=}; \textit{Last accessed on March 19th, 2025.}}, the zbMath API\footnote{\url{https://api.zbmath.org/v1/document/}; \textit{Last accessed on March 19th, 2025.}}, and the ORCiD API\footnote{\url{https://pub.orcid.org/v3.0}; \textit{Last accessed on March 19th, 2025.}} for the complete metadata of the publication and its authors.

Next to the documentation catalogs, MaRDMO provides a dedicated search catalog that allows users to explore existing entities. Users can search for software, methods, or mathematical models and discover interdisciplinary workflows that utilize these entities. Similarly, mathematical models can be retrieved based on their application in specific tasks or their role in modeling research problems. Algorithms can be found based on the problems they solve or the software in which they are implemented. Additionally, keyword-based searches enable users to explore interdisciplinary workflows via research objectives, models via research problems, and algorithms via algorithmic problems. This search functionality provides an accessible interface for querying the individual MaRDI services while abstracting away the complexity of SPARQL.

\subsection{MaRDMO Export}\label{sec:mardmo-export}

The MaRDMO Plugin integrates into the RDMO interface by adding a ``MaRDMO Button'' to the export section. This button enables researchers to export their documentation but is only functional if one of the MaRDMO catalogs has been used.

The export process follows a structured workflow. Once a documentation process is completed—or search parameters are selected—the export function first provides a preview. This preview presents the collected information in the standardized documentation template for interdisciplinary workflows, mathematical models, or algorithms, depending on the type of documentation. In the case of the search catalog, the SPARQL query generated by MaRDMO is displayed, allowing users to inspect the exact query executed on their behalf. While generating the preview, relevant publication information is retrieved if DOIs are provided. The thus obtained citation information is stored in the background for the final export.

If the documentation is considered to be correct, the user can proceed by clicking ``Export to MaRDI Service,'' which triggers the final export. The export mechanism differs based on the type of research data. For mathematical models and algorithms, MaRDMO generates SPARQL INSERT queries to add the new information into the MathModDB or MathAlgoDB knowledge graphs. For interdisciplinary workflows, the export is handled via the REST API of the MaRDI Portal. This process follows an OAuth2 authentication protocol, allowing users to log in to the MaRDI Portal using their institutional credentials. The workflow documentation is then added to the MaRDI Portal under the user's name. While MaRDMO adds machine-readable information\footnote{\url{https://portal.mardi4nfdi.de/wiki/Item:Q6032641}; \textit{Last accessed on March 25th, 2025.}} to the knowledge graphs, human-readable representations\footnote{\url{https://portal.mardi4nfdi.de/wiki/Workflow:6032641}; \textit{Last accessed on March 25th, 2025.}} are automatically generated through structured templates\footnote{\url{https://portal.mardi4nfdi.de/wiki/Template:Workflow}; \textit{Last accessed on March 25th, 2025.}} on each platform. 

To maintain a certain level of data quality, a two-step validation process is envisioned. First, MaRDMO ensures automated ontological validation before export, verifying that all relationships follow the structure of the respective ontologies (MathModDB, MathAlgoDB, and interdisciplinary workflow) and checking whether the documentation is complete and connected. A documentation is considered complete and connected if it establishes all necessary relationships. However, users are not required to provide all related information themselves. For example, if a user only wants to add a new computational task, it must be ensured that the task is connected to a mathematical model, mathematical formulations and quantities as indicated in the automating scheme (c.f. Fig. \ref{fig:figure2}). However, connected and complete does not mean that relations within the individual classes need to be established. The second step of validation is a review by domain experts, which is carried out by the individual knowledge graph providers to ensure the accuracy and relevance of the content.

Once the export is completed, MaRDMO presents a landing page with direct links to the newly created items in the respective knowledge graphs. Additionally, MaRDMO updates the filled-out questionnaire. For newly created items in either knowledge graph, MaRDMO automatically appends the persistent identifiers (PIDs) to the questionnaire for future reference. For search queries, MaRDMO provides a summary of the SPARQL results, allowing a direct access of the relevant workflows, models, or algorithms.

\section{Future Potential and Developments}\label{sec:future-potential-and-developments}

The MaRDMO Plugin has the potential to significantly enhance the dissemination of mathematical research data across disciplines. Traditionally, the transfer of mathematical models and algorithms into new fields relies heavily on personal connections and interdisciplinary collaborations. While these collaborations are valuable, they also introduce limitations—especially when researchers are unaware of potential applications of their work in other domains.

For example, consider a mathematician who develops a model and an algorithm to extract logical rules from a dataset by transforming (data) objects into Boolean polynomials and computing the Gröbner basis of the corresponding ideal. If this work is only published in a mathematics journal, it is unlikely to reach researchers in fields such as Egyptology, where it might be applicable for analyzing destruction patterns in ancient egyptian statues. Without an accessible and structured way to describe and search for such models, their interdisciplinary use remains restricted. MaRDMO addresses this challenge by enabling researchers to search for models and algorithms based on their functionality—rather than requiring detailed knowledge of mathematical theory in the first place. By enabling structured search, MaRDMO helps researchers efficiently locate concise descriptions, relevant metadata, and implementation resources—provided by individual MaRDI services—thereby lowering the barrier for non-mathematicians to explore and adopt mathematical research data. Additionally, researchers from non-mathematical disciplines can document their own interdisciplinary workflows—such as applying logical data analysis to ancient egyptian objects—allowing these workflows to be discovered and reused in new contexts, such as medical research.

Beyond facilitating discoverability, MaRDMO also enhances research data management through PIDs. When researchers document their models, algorithms, or workflows in MaRDI services, they receive a unique PID, which can be referenced in publications. This improves the findability of research outputs and provides a quality assurance mechanism, as data integrated into MaRDI services follows structured documentation guidelines and validation. Such PIDs can also be added to DMPs to enhance the data provenance section.

Looking ahead, MaRDMO will expand to support additional MaRDI services, such as MaRDIFlow~\cite{Veluvali2023}. A key focus will also be on integrating additional sources to further enrich interdisciplinary documentation. For example, when documenting a workflow which includes experimental measurements through a specific measurement device, MaRDMO will retrieve relevant metadata from databases maintained by other NFDI consortia or domain-specific repositories\footnote{\url{https://o2a-data.de/}; \textit{Last accessed on March 20th, 2025.}}. This integration will be facilitated in collaboration with the DMP4NFDI basic service, allowing researchers to seamlessly incorporate external data sources into their documentation.

Another ongoing initiative is the development of a dedicated MaRDI RDMO instance, in collaboration with DMP4NFDI. This will provide researchers from all disciplines with direct access to MaRDMO, even if their institution does not offer an RDMO instance.

Finally, MaRDMO is exploring the potential use of Artificial Intelligence (AI) to assist in the documentation process. One potential approach is to use AI to analyze research papers and map their content onto existing ontologies, automatically identifying relevant aspects of mathematical models, algorithms, and interdisciplinary workflows. This could significantly reduce the manual effort required for documentation and improve the integration of research outputs into MaRDI services. AI-driven techniques could also help establish new connections between mathematical models and potential application areas, further supporting the interdisciplinary diffusion of mathematical research data.

\section*{Acknowledgements}
M. Reidelbach is supported by MaRDI, funded by the Deutsche Forschungsgemeinschaft (DFG), project number
460135501, NFDI 29/1 “MaRDI – Mathematische Forschungsdateninitiative”.

\section*{Authorship Contributions}
M. Reidelbach contributed to the conceptualization, tool development, and writing (first draft, review and editing). 

\section*{Conflict of Interest}
The author declares that there are no conflicts of interest.
\newline\newline
\bibliographystyle{plain}
\bibliography{literature}

\end{document}